\documentclass[conference]{IEEEtran}
\IEEEoverridecommandlockouts
\usepackage[pdftex]{graphicx}
\usepackage{amsfonts}
\usepackage[cmex10]{amsmath}
\usepackage{mathabx}
\usepackage{algorithmic}
\usepackage{cite}
\usepackage{array}
\usepackage{mdwmath}
\usepackage{mdwtab}
\usepackage{eqparbox}
\usepackage{url}
\usepackage{float}
\usepackage{booktabs,chemformula}
\usepackage{hyperref}
\usepackage{cleveref}
\usepackage{fancyhdr}
\usepackage[super]{nth}
\usepackage[pscoord]{eso-pic}
\usepackage{amssymb}

\def\BibTeX{{\rm B\kern-.05em{\sc i\kern-.025em b}\kern-.08em
    T\kern-.1667em\lower.7ex\hbox{E}\kern-.125emX}}
\begin{document}

\title{DSE Stock Price Prediction using Hidden Markov Model}

\author{\IEEEauthorblockN{Raihan Tanvir\IEEEauthorrefmark{1}\textsuperscript{\textsection}, Md Tanvir Rouf Shawon\IEEEauthorrefmark{1}\textsuperscript{\textsection}, Md. Golam Rabiul Alam\IEEEauthorrefmark{2}}
\IEEEauthorblockA{\IEEEauthorrefmark{1}Ahsanullah University of Science and Technology, Dhaka, Bangladesh}
\IEEEauthorblockA{\IEEEauthorrefmark{2}Brac University, Dhaka, Bangladesh}
\IEEEauthorblockA{\{raihantanvir.96, shawontanvir95\}@gmail.com, rabiul.alam@bracu.ac.bd}
}
\maketitle
\begingroup\renewcommand\thefootnote{\textsection}
\footnotetext{Authors have equal contribution}
\endgroup

\begin{abstract}
Stock market forecasting is a classic problem that has been thoroughly investigated using machine learning and artificial neural network based tools and techniques. Interesting aspects of this problem include its time reliance as well as its volatility and other complex relationships. To combine them, hidden markov models (HMMs) have been utilized to anticipate the price of stocks. We demonstrated the Maximum A Posteriori (MAP) HMM method for predicting stock prices for the next day based on previous data. An HMM is trained by analyzing the fractional change in the stock price as well as the intraday high and low values. It is then utilized to produce a MAP estimate across all possible stock prices for the next day. The approach demonstrated in our work is quite generalized and can be used to predict the stock price for any company, given that the HMM is trained on the dataset of that company's stocks dataset. We evaluated the accuracy of our models using some extensively used accuracy metrics for regression problems and came up with a satisfactory outcome.
\end{abstract}


\section{Introduction}
Stock price forecasting has been one of the most difficult issues for the AI community. Typical AI research, which is primarily focused on building intelligent solutions that are meant to imitate human intelligence, has typically gone beyond the limits of forecasting research. Stock price forecasting, however, remains severely constrained because of its non-stationary, cyclic, and stochastic character. A variety of factors influence the rate of price fluctuations in such a series, including equity, the rate of interest, options, securities, warrants, mergers and acquisitions of significant financial organizations, and so on. In such market, ordinary investors can not profit regularly. As a result, an intelligent forecasting model for the stock market would be highly demanded and of significant interest to ordinary investors.

HMMs are seen as successful in analyzing and forecasting time-dependent events. This technique has already been used for voice recognition \cite{speech}, handwriting recognition \cite{handwritten}, facial expression recognition \cite{facial}, ECG analysis \cite{ecg}, and other purposes. Stock market forecasting is analogous to these problems in terms of its inherent relationship with time. Based on an unseen collection of states from which transitions can be made, hidden Markov models correlate each state with a probable observation. In a similar fashion, the stock market can be seen. Investors are usually unaware of the inherent factors that influence share prices. Transitions between these underlying states are influenced by business strategy, decisions, the market environment, etc. The stock's value is the observable result that reflects these. So, HMM obviously complies with this real-world setting.

The selection of features is very crucial in this method. Several attempts have been made in the past to use the volume of trade, the stock's momentum, as well as the market's correlation and volatility. We incorporate the daily fractional differences in stock values as well as the fractional deviation of intraday maximum and minimum values, as proposed in \cite{base-1}. It is essential to comprehend the fractional change to generate the desired prediction. The fractional difference between the intraday high and low values is a good predictor of volatility direction.

Despite the fact that HMMs have been utilized in this area for a long time, none of the works have contributed to the Bangladesh stock market. As a response, we opted to use a HMM-based approach to serve this purpose. To employ the technique, we use stock data from companies listed on the Dhaka Stock Exchange (DSE). For each stock, a unique HMM is trained. The sole constraint that the training dataset must satisfy is significant variability in the observations. It is resolved by effectively using long spans of time (13 years) during which the stock price swings consistently, albeit significantly.

\section{Previous Works}
Various studies have been conducted recently in an attempt to develop a stock market forecasting model that is flawless (or close so). In most of the forecasting research, statistical time series analysis methodologies such as the auto-regression moving average (ARMA) ~\cite{arma} and multiple regression approaches are utilized.

A HMM based approach for stock price forecasting \cite{base-1} is deeply studied. For predicting the following day\'s stock value given historical data, the authors followed the Maximum a Posteriori HMM method. The continuous HMM is trained using the fractional difference in stock values and the stock's intraday highs and lows. After much deliberation, we decided to use the technique outlined in this paper to develop our model.

In \cite{svm-based}, a directed-weighted chunking SVMs approach is described, where the complete training dataset is partitioned into numerous sections and support vectors for each portion are created. To create the forecast model, weighted support vector regressions are computed on the new working data set.

Md. Rafiul Hassan et al. introduced ~\cite{base-2}, an HMM-based model in which predictions are generated by interpolating the adjacent values of the dataset. The results achieved from this experiment are inspiring, and a novel framework for stock market analysis is explored.

Using autonomously generated fuzzy connections, Romahi Y et al. introduced a unique technique to dynamic financial forecasting \cite{fuzzy-based}. This method has yielded promising results, but building a fuzzy system requires domain expertise.

H. Liu et al. offered a deep residual network based strategy for prediction where the stock price graph is utilized as an input \cite{resnet}. This model's average accuracy was $0.40$, which is higher than the stochastic indicator's average accuracy of $0.33$.

Also, a naive approach demonstrated in the official documentation of the \emph{hmmlearn} package is thoroughly investigated. In this scheme, only the difference between two consecutive closing prices of a stock is considered. Though it's a simple approach and ignores several key factors in stock price prediction, it has shown quite satisfactory results.

\section{Methodology}
An HMM, $\lambda$ may be represented as,

$\lambda=(\pi, A, B)$,

where $A$ is the transition matrix, the entries of which represent the likelihood of a state switching from one state to another, $B$ is the emission matrix, that provides $b_{j}\left(O_{t}\right)$ the probability of witnessing $O_{t}$ while in state $j$ and, $\pi$ represents the initial probabilities of the states at time, $t=1$. We consider the emission probability distribution to be continuous, since the samples are a vector of continuous random variables. For simplicity, we'll consider it a multinomial Gaussian distribution having parameters, ($\mu$ and $\Sigma$)

\begin{figure}
	\centering
	\includegraphics[width=\columnwidth]{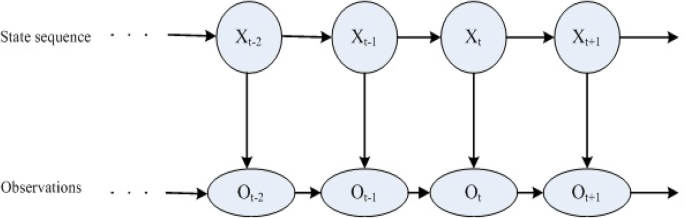}
	\caption{Hidden Markov Model} 
	\label{fig:hmm_basic}
\end{figure}

%
%where  $A$ is the transition matrix whose elements give the probability of a transition from one state to another, $B$ is the emission matrix giving $b_{j}\left(O_{t}\right)$ the probability of observing $O_{t}$ when in state $j$, and  $\pi$ gives the initial probabilities of the states at $t=1$. Further, as the observations are a vector of continuous random variables, assume that the emission probability distribution is continuous.%
%
%For simplicity, we'll consider it's a multinomial Gaussian distribution with parameters ($mu$ and $Sigma$). So, with the transition matrix, A, prior probabilities, $Pi$, as well as $mu$ and $Sigma$, which represent the multinomial Gaussian distribution, we must determine the following parameters.%
\iffalse
\begin{equation}
	b_{j}\left(\overrightarrow{O_{t}}\right)=\sum_{m=1}^{M} c_{j m} N\left(\overrightarrow{O_{t}}, \overrightarrow{\mu_{j m}}, \Sigma_{j m}\right)
\end{equation}
where:
\begin{itemize}
	\item $M$ is the number of Gaussian Mixture components.
	\item $c_{j m}$ is the weight of the mixture component in state $j$.
	\item $\overline{\mu_{J m}}$ is the mean vector for the $m^{th}$ component in the state $j^{th}$
	\item $N\left(\overrightarrow{O_{t}}, \overrightarrow{\mu_{J m}}, \Sigma_{j m}\right)$ is the probability of observing $\overline{O_{t}}$ in the multi-dimensional Gaussian distribution.
\end{itemize}\fi
%
%The model's observations comprise daily stock data in the form of a vector having three entries,
The model's observation is a three dimensional vector representing daily stock information,
\begin{equation}
	\label{eq:obs}
	\begin{aligned}
		O_{t}=&\left(\frac{\text { close }-\text { open }}{\text { open }}, \frac{\text { high }-\text { open }}{\text { open }}, \frac{\text { open }-\text { low }}{\text { open }}\right) \\
		&:=(\text { $frac_{Change}, frac_{high}, frac_{Low}$ })
	\end{aligned}
\end{equation}
In ~\eqref{eq:obs} open represents the day's opening price, close denotes the day's closing price, high and low represent the day's highest price and lowest price, respectively. To characterize the variance in stock values that stays constant throughout time, we utilize fractional changes.

After training, an approximation of the Maximum a Posteriori (MAP) technique is employed to test it. When projecting future stock prices, we assume a $d$ day latency. As a result, having the HMM model, $\lambda$ and the stock prices for previous $d$ days $\left(O_{1}, O_{2}, \ldots, O_{d}\right)$, as well as the opening price for the $(d+1)^{st}$ day, the task is to calculate the closing price for the $(d+1)^{st}$ day, which is equivalent to computing the fractional difference for the $(d+1)^{st}$ day, $\frac{close - open}{open}$. This is done using the MAP approximation of the observation vector, $O_{d+1}$.
%
%Once the model is trained, testing is done using an approximate Maximum a Posteriori (MAP) approach. We assume a latency of $d$ days while forecasting future stock values. Hence, the problem becomes as follows - given the HMM model $\lambda$ and the stock values for $d$ days $\left(O_{1}, O_{2}, \ldots, O_{d}\right)$ along with the stock open value for the $(d+1)^{st}$ day, we need to compute the close value for the $(d+1)^{st}$ day. This is equivalent to estimating the fractional change $\frac{close - open}{open}$ for the $(d+1)^{st}$ day. For this, we compute the MAP estimate of the observation vector $\left(O_{d+1}\right)$.%

Lets consider $\hat{O}_{d+1}$, the MAP value of the observation on the $d+1$ day, provided the values of the previous $d$ days.
\begin{equation}
	\hat{O}_{d+1}=\arg \max _{o_{d+1}} P\left(O_{d+1} \mid O_{1}, O_{2}, \ldots, O_{d}, \lambda\right)
\end{equation}
\begin{equation}
	=\arg \max _{o_{d+1}} \frac{P\left(O_{1}, O_{2}, \ldots, O_{d}, O_{d+1} \mid \lambda\right)}{P\left(O_{1}, O_{2}, \ldots, O_{d}, \lambda\right)}
\end{equation}
The observation vector  $\left(O_{d+1}\right)$ is adjusted throughout the whole range of potential values. Because the denominator is invariant with regard to $\left(O_{d+1}\right)$ the MAP approximation reduces to ~\eqref{eq:reduced_map}.
\begin{equation}
	\label{eq:reduced_map}
	\hat{O}_{d+1} = \arg \max _{o_{d+1}} P\left(O_{1}, O_{2}, \ldots, O_{d}, O_{d+1} \mid \lambda\right)
\end{equation}
We determine the maximum probability by computing the probability across a distinct set of potential $O_{d+1}$ values. The computational cost of determining the likelihood of a particular observation is $O(n^2 d)$, where $n$ denotes the number of states and $d$ denotes the latency. This process is repeated for all distinct set of potential values of $O_{d+1}$. We have $n = 4, d = 30$ and the number of possible values of $O_{d+1}$ is $50 \times 10 \times 10$ (see table \ref{table:map-values}).
\begin{figure}[htb]
	\centering
	\includegraphics[width=\columnwidth, height=\columnwidth]{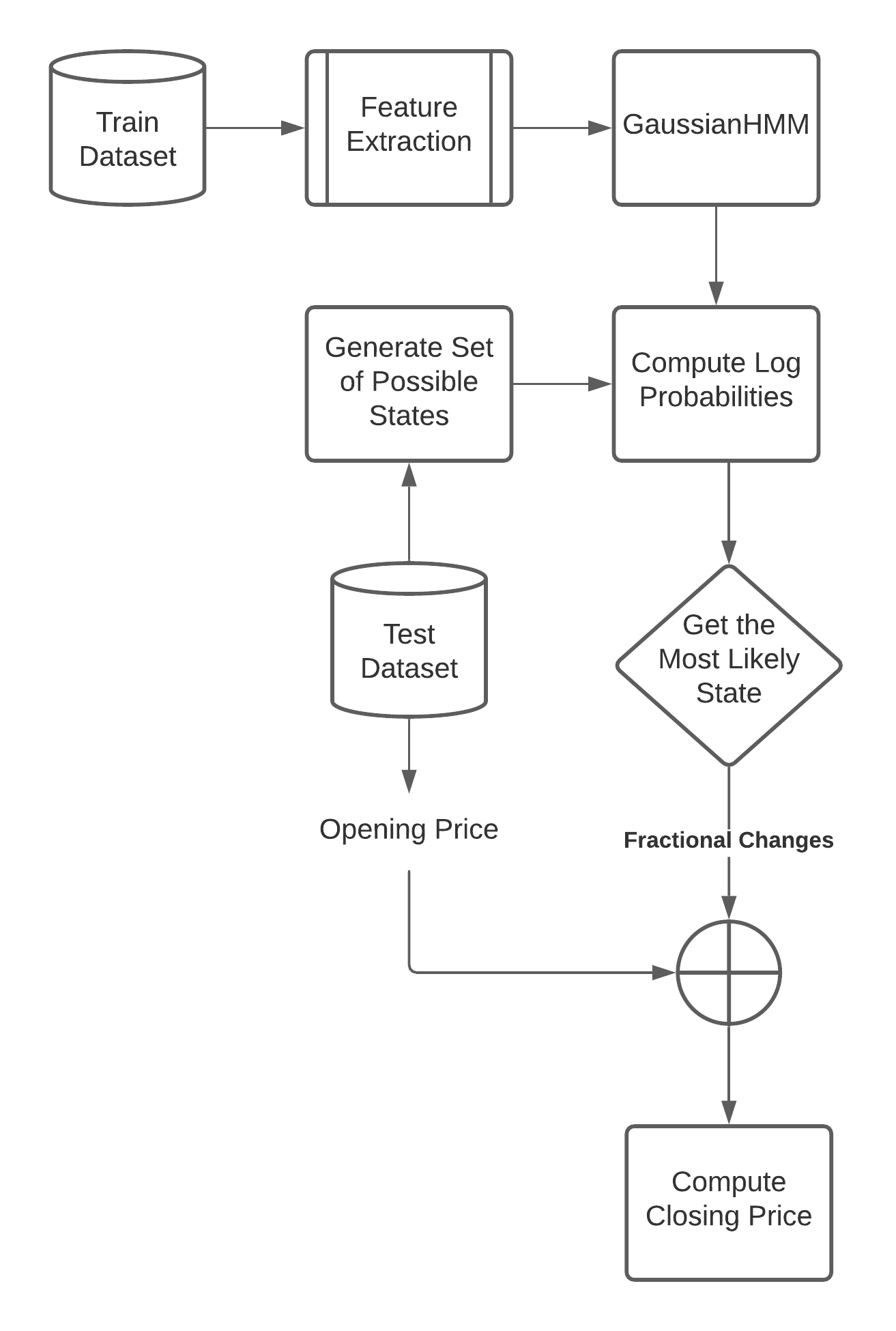}
	\caption{Predicting Closing Price Using Fractional Changes}
	\label{fig:block_diag}
\end{figure}
The closing price of a given day can be determined by taking the day's opening price and incorporating it with the expected fractional difference for that day as shown in ~\eqref{eq:close_p}.
\begin{equation}
	closing\_price = opening\_price \times (1 + frac_{change})
	\label{eq:close_p}
\end{equation}
We also employed a naive approach, where only the differences in the closing prices of two successive days of stock are considered. The observations in this approach are the difference in closing prices for two consecutive days and the volume of stocks. After training the model, a prediction is made by computing the inner product of the transition matrix and the mean of the observations. Finally, the predicted changes are incorporated with the previous day's closing price to generate the ultimate prediction.

\section{Dataset}
To put the chosen technique into practice, we selected the DSEBD\footnote{\url{https://www.kaggle.com/mahmudulhaque/dsebd}\label{DSEBD}} dataset from \href{https://www.kaggle.com}{Kaggle}. The dataset consists of annual stock data of companies registered to Dhaka Stock Exchange (DSE) from January 2008 to December 2020 in separate JSON files. The dataset contains stock records for a total of $589$ companies from $22$ different sectors with $5$ unique instrument types. The number of observations in the dataset is $15,75,134$.

To fit the dataset into our scheme, the following steps of processing have been performed.
\begin{itemize}
	\item The json files are loaded into pandas data-frame and stored into a list of data-frames
	\item The items of the list are concatenated into a single data-frame
	\item The concatenated frame is saved as csv.
	\item Separate Dataset is created by extracting the data for individual stocks from the concatenated frame.
\end{itemize}
\iffalse
\begin{table}
	\caption{Various Statistics of DSEBD Dataset}
	\label{table:dsebd}
	\centering
	\begin{tabular}{|c|c|c|l|}
		\hline
		%\multicolumn{4}{|c|}{DSEBD Dataset}     \\ \hline
		Company & Sector & Type & Records \\ \hline
		589     &  22     & 55   & 1575134 \\ \hline
	\end{tabular}%
\end{table}\fi
%
%When the training and test datasets are separated, the data is less likely to be over-fitted into the model. This is why the dataset is broken down into two categories: the train data for training the model and the test set are used to deliver an objective assessment of the final model fit on the training dataset. Using the \emph{train\_test\_split} function supplied by the sklearn.model selection module, we have accomplished this goal.%
%
\begin{figure}
	\centering
	\includegraphics[width=1\columnwidth]{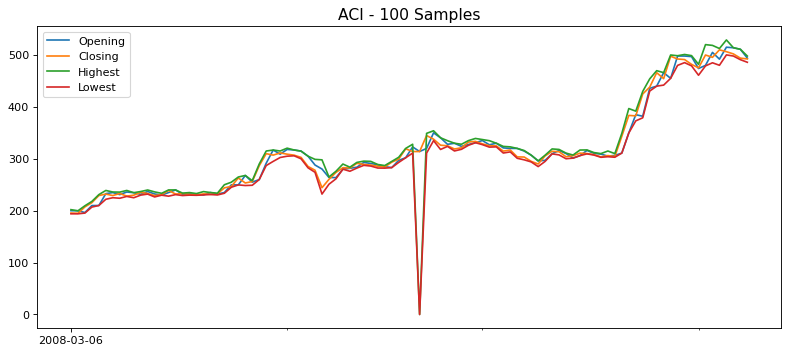}
	\caption{Trend of opening, closing, highest and lowest prices for ACI stocks (100 samples)} 
	\label{fig:100_samples}
\end{figure}
The dataset is split into $0.8 : 0.2$ ratio using \emph{train\_test\_split} function supplied by sklearn.model\_selection module, where $80\%$ data is used in train set and rest of the $20\%$ is used for test set. As we are working with time-series data, we need to preserve the temporal sequence of the data. To avoid random splitting of data into train and test sets, we passed \emph{shuffle=False} as the parameter.

\begin{figure}
	\centering
	\includegraphics[width=1\columnwidth]{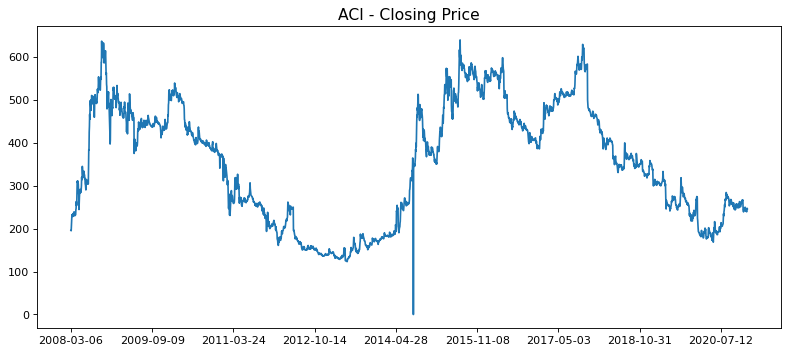}
	\caption{Trend of closing prices for ACI's Stocks from Jan 2008 to Dec 2020.} 
	\label{fig:closing_p_aci}
\end{figure}

The trend of the stock price of \textbf{ACI} Limited is shown in the figure \ref{fig:100_samples} and \ref{fig:closing_p_aci}. Table \ref{table:aci_df} shows some sample data from \textbf{ACI} stocks dataset.
\begin{table}
	\centering
	\caption{Sample data from ACI stocks dataset}
	\label{table:aci_df}
	\begin{tabular}{@{}ccccccc@{}}
		\toprule
		\textbf{date} & \textbf{open} & \textbf{high} & \textbf{low} & \textbf{close} & \textbf{volume} & \textbf{prev\_close} \\ \midrule
		2008-03-06 & 200.0 & 202.0 & 194.0 & 195.5 & 266850 & 198.8 \\
		2008-03-09 & 199.8 & 199.8 & 194.0 & 195.0 & 333600 & 195.5 \\
		2008-03-09 & 199.8 & 199.8 & 194.0 & 195.0 & 333600 & 195.5 \\
		2008-03-10 & 196.5 & 209.5 & 195.4 & 207.3 & 381650 & 195.0 \\
		2008-03-11 & 209.9 & 217.9 & 207.0 & 215.5 & 509550 & 207.3 \\ \bottomrule
	\end{tabular}%
\end{table}
\section{Implementation}
As our HMM, we employed the GaussianHMM class from the hmmlearn\footnote{\url{https://hmmlearn.readthedocs.io/en/latest/}\label{hmmlearn}} package and performed parameter approximation using the \emph{fit} method provided by it. Because we investigated two approaches, we'll go through the implementation specifics of the first technique, which uses fractional changes in stock prices. Following that, the execution of a later strategy that takes into account successive price fluctuations in stocks is addressed.
\subsection{HMM with Fractional Changes}
\subsubsection{Initialization:} The initialization of the HMM is done by setting the parameters according to following configurations:
\begin{itemize}
	\item Quantity of hidden states,  $n$ = 4
	\item Dimension of observations, $D$ = 3
	\item Latency,  $d$ = 30 days
	\item Max number of iterations, $n\_iter = 10000$
	\item Convergence threshold $tol = 0.001$
\end{itemize}
These values are obtained from \cite{base-1}, on the basis of which we're developing our model. Furthermore, ~\cite{state-n} recommended using $4$ underlying states because the dimension of observation is likewise $4$. The remaining model parameters are initialized with the default values of the GaussianHMM class provided by the \emph{hmmlearn} package.
\subsubsection{Training:}
\begin{figure}
	\centering
	\includegraphics[width=\columnwidth]{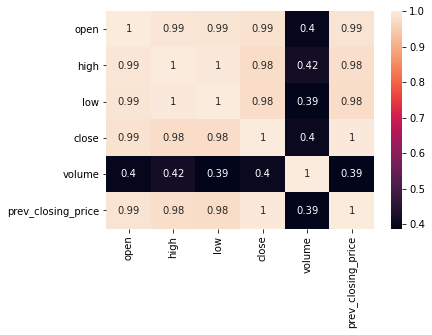}
	\caption{Correlation matrix for ACI stocks data} 
	\label{fig:corr_mat}
\end{figure}

Though our dataset has many features, not all of them have a strong influence on the closing price of a stock. From the correlation matrix in figure \ref{fig:corr_mat}, we can see that the opening price, highest price, and lowest price have a strongly positive relationship with the closing price (exactly 1 or very close to 1). Hence, we consider these four features to be the attributes of our interest.

Now, we have relatively few characteristics for each day, primarily the starting and ending prices of the stock for that day, as well as the maximum and minimum prices of the stock. So, we utilize them to compute stock prices. Instead of directly using these values, we extracted the fractional differences in each of them that would be used to train our HMM. 
%To serve this purpose a feature extraction method is used as shown in code snippet ~\ref{listing:feature_extraction}.
%

Following feature extraction, some rows may include NaN, infinity, or values that are too big for dtype('float64'). A data cleansing procedure is carried out to remove such data.
% as shown in code snippet ~\ref{listing:data_cleansing}.
%
After performing the aforementioned two operations, the HMM is trained using the \emph{fit} method of \textbf{GaussianHMM} class with the obtained feature vectors. Though the $n\_iter$, was set $10000$, yet the training was early stopped after $35$ iterations, due to the convergence threshold, $tol$.
\begin{figure*}[htb]
	\centering
	\includegraphics[width=0.75\linewidth]{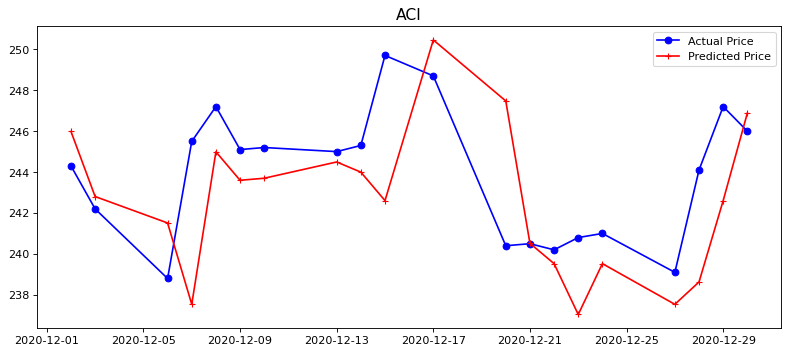}
	\caption{Actual v/s Forecasted Value for ACI stocks from Dec 02, 2020 to Dec 30, 2020 by  HMM with Fractional Changes.} 
	\label{fig:pred_aci}
\end{figure*}
\begin{figure*}[htb]
	\centering
	\includegraphics[width=.75\linewidth]{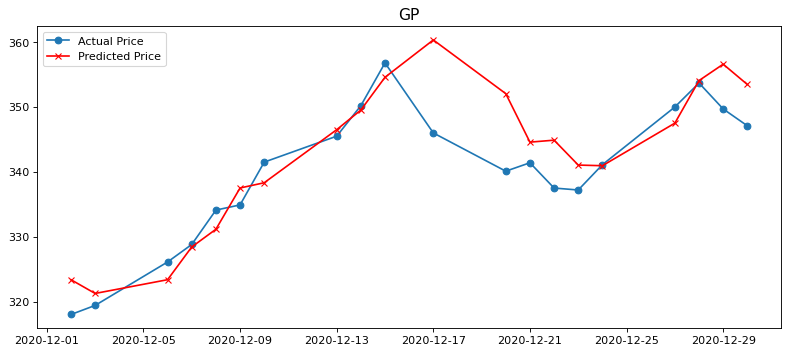}
	\caption{Actual v/s Forecasted Value for GrameenPhone stocks from Dec 02, 2020 to Dec 30, 2020 by HMM with Fractional Changes.}
	\label{fig:pred_gp}
\end{figure*}
\subsubsection{Prediction:}
Once our model has been trained, we then estimate the stock's closing price.  Provided the opening stock price for a day and the data from the previous $d$ days, we may compute the closing stock price for that day. Our predictor would have a latency of $d$ day. This means that if we can estimate $frac_{change}$ for a particular day, we can calculate the closing price using the \eqref{eq:close_p}.

The optimization of the problem would have been computationally expensive if $frac_{change}$ is considered as a continuous variable. As a result, we discretized them into values within the boundaries of two finite variables (as shown in the table ~\ref{table:map-values}) and generated a set of fractional changes, $<frac_{change} , frac_{high}, frac_{low}>$ by deriving the cartesian products of these three variables. A higher number of steps is used for the $frac_{change}$ since these are eventually utilized for stock prediction.
\begin{table}
	\caption{Range of values for dividing fractional changes}
	\label{table:map-values}
	\centering
	\begin{tabular}{@{}cccc@{}}
		\toprule
		\textbf{Observation} & \textbf{Min Value} & \textbf{Max Value} & \textbf{Number of Steps} \\ \midrule
		$frac_{change}$ & -0.1 & 0.1 & 50 \\ \midrule
		$frac_{high}$   & 0    & 0.1 & 10 \\ \midrule
		$frac_{low}$    & 0    & 0.1 & 10 \\ \bottomrule
	\end{tabular}%
\end{table}
Then, the log probability under the model for the observation sets from the previous $d$ days are derived using the \emph{score} method of the GaussianHMM class, and the maximum is found. This way, the most probable outcome i.e: $frac_{change}$ is found and the closing price is computed using \autoref{eq:close_p}. For predicting $d$ days stock price, this method is called iteratively for those days index. Figure \ref{fig:pred_aci} and ~\ref{fig:pred_gp} shows the actual stock price along with the forecasted price using our trained model for \textbf{ACI Limited} and \textbf{Grameenphone Limited} respectively.
The method we are following is a generalized approach and can be used to build a predictor for any company given that the HMM is trained on the dataset of that company's stock. Hence, we could use the same approach and build a separate prediction model for each of the companies listed on the Dhaka Stock Exchange.
\subsection{HMM with Successive Fluctuations}
We followed the same settings for initialization in this method as in the previously described approach. The difference is in the feature set considered for this technique.
\begin{figure}[H]
	\centering
	\includegraphics[width=\columnwidth]{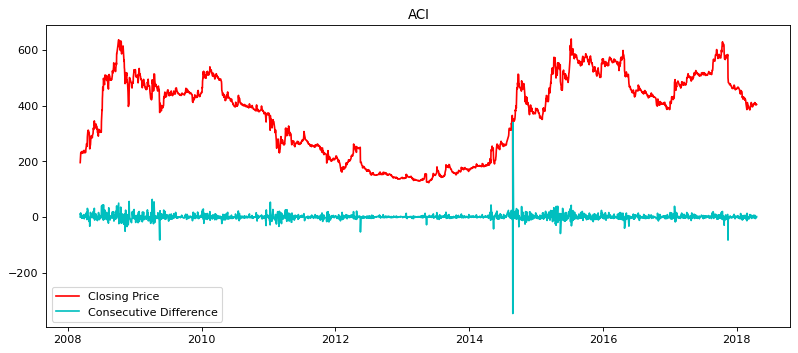}
	\caption{The fluctuations in closing prices for ACI stocks.} 
	\label{fig:consec_diff}
\end{figure}
\subsubsection{Training:}
To train the model with this approach, we first extracted the closing prices and volumes of the stocks. Then the successive difference between stock prices is calculated and stored. Finally, the tuple of dates, consecutive differences, and volumes is stacked vertically to form the feature vectors. Finally, the model is trained with these feature vectors. It took $240$ iterations for the convergence of the model.
\subsubsection{Prediction:}
\begin{figure*}
	\centering
	\includegraphics[width=.75\linewidth]{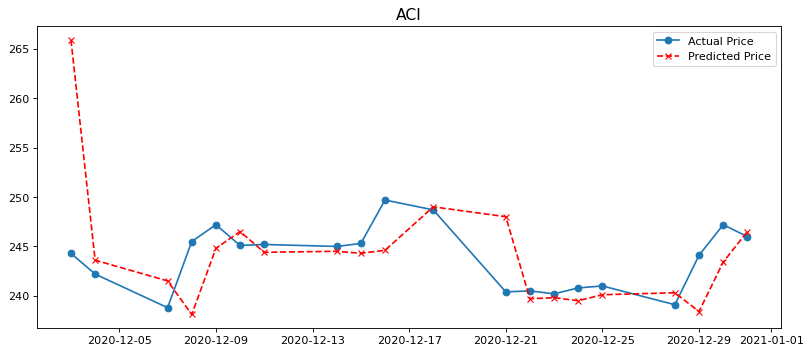}
	\caption{Actual v/s Forecasted Value for \emph{ACI Ltd}'s from Dec 02, 2020 to Dec 30, 2020 by HMM with Successive Fluctuations.} 
	\label{fig:pred_2}
\end{figure*}

Since our model is trained to detect the pattern of changes in the closing price, it is ready to make a prediction of the changes that would appear in the closing price of the next day, given the closing price of the previous day. To achieve this objective, the dot product of the transition matrix $A$ and the mean of the data distribution are computed. Then we incorporated the value of the expected change with the value of the previous closing price to make our final predictions on the closing price of that day. Figure \ref{fig:pred_2} illustrates the \emph{ACI Limited's} actual stock price along with the predicted price.
\subsection{Evaluation}
To assess the performance of our models, three widely accepted performance metrics for regression task have been used.

\textbf{Mean Absolute Error (MAE)} is a straightforward measure that computes the absolute difference between actual and projected values. It is the most robust metric for outliers.

\textbf{Root Mean Squared Error (RMSE)} is self-explanatory; it is the square root of the mean squared error(MSE). MSE is the mean of squared difference between the true and predicted values.

\textbf{Mean Absolute Percentage Error (MAPE)} is the mean absolute error between the true and forecasted values in percentage.
\begin{equation}
	MAPE=\frac{1}{n} \sum_{i=1}^{n} \frac{\left|p_{i}-y_{i}\right|}{\left|y_{i}\right|} \times 100 \%
\end{equation}
here $y_{i}$ and $p_{i}$ is the true and forecasted values respectively and $n$ is the number of days for which the data are evaluated. 

Table ~\ref{table:pred} shows the metric scores for the ACI Limited\'s stocks using our trained model. Here we can see that, the \emph{MAPE} values for the models' are $1.0265$ and $1.3672$. %As stated in  a MAPE score of less than $5$ suggests that the forecast is acceptable in terms of accuracy.
A MAPE score of less than $5$ indicates that the forecast's accuracy is acceptable, according to \cite{swanson}. So we can conclude that our predictors' performance is quite excellent, leaving the limitations described in the later section aside.
\begin{table}
	\centering
	\caption{Evaluation of predictions made on ACI stocks}
	\label{table:pred}
	\begin{tabular}{ccc}
		\hline
		\textbf{Metric} & \textbf{$HMM_{frac_{change}}$} & \textbf{$HMM_{succ_{fluctuation}}$} \\ \hline
		MAE             & 2.5064                & 3.3382                \\ \hline
		RMSE            & 3.4003                  & 5.8141               \\ \hline
		MAPE            & 1.0265                  & 1.3672                \\ \hline
	\end{tabular}%
\end{table}
\section{Limitations and Future Works}
We have considered a dataset that contains stock data from January 2008 to December 2020. There are many open-source scrapping tools that facilitate the retrieval of the upto date stock information. Due to time constraints, we could not employ them. The stock market has a volatile property, and the prices of stocks may fluctuate to a large extent. For this type of scenario, our models fail to provide accurate predictions. We are willing to overcome this limitation by using various smoothing techniques in the future.

\section{Conclusion}
Artificial intelligence and machine learning techniques have been frequently used to forecast stock values in Bangladesh's stock market. To the best of our knowledge, none of them have yet exploited the efficiency of hidden markov models in their work. In our paper, we have demonstrated the capability of HMMs to predict stock prices using historical data. Two approaches have been investigated. And both of them provide satisfactory outcomes for typical cases. If we can gather more domain knowledge and apply techniques for handling the abrupt changes in stocks' values, then we may be able to achieve perfection in the accuracy of our model. Finally, we believe that our work will extend the window of research on forecasting using the Hidden Markov Model.

	 \bibliographystyle{IEEEtran}
	 \bibliography{ref}

\end{document}